\documentstyle[12pt]{article}
\begin{document}
\title{COSMIC BACKGROUND RADIATION DUE TO PHOTON CONDENSATION}
\author{B.G. Sidharth$^*$\\ Centre for Applicable Mathematics \& Computer Sciences\\
B.M. Birla Science Centre, Hyderabad 500 063 (India)}
\date{}
\maketitle
\footnotetext{E-mail:birlasc@hd1.vsnl.net.in}
\begin{abstract}
It is shown that a collection of photons with nearly the same frequency
exhibits a Bose "condensation" type of phenomenon at about $3^\circ K$ corresponding
to a peak intensity at a wave length of about $0.4cm.$ This could give a
mechanism for the observed Cosmic Background Radiation, and also explain
some curious features.
\end{abstract}
It is a curious circumstance that the substitution of the cosmic microwave
background radiation photon number density in the formula for the Bose-Einstein
condensation of a relativistic gas gives a temperature of around $3^\circ K$ or
Bose-Einstein condensation of the cosmic photon radiation takes place at
about this temperature\cite{r1}. In this light we consider the following scenario:
An assembly of photons, most of them having nearly the same frequency in or nearly
in equilibrium. We now show that at a temperature of about $3^\circ K$, a
condensation takes place or conversely, corresponding to a microwave
radiation of about $4mm$. The interesting thing is that this derivation is
independent of any cosmological model.\\
Our starting point is the formula for the average occupation number for photons
of momentum $\vec k$ for all polarizations\cite{r2}:
\begin{equation}
\langle n_{\vec k} \rangle = \frac{2}{e^{\beta \hbar \omega}- 1}\label{e1}
\end{equation}
Let us specialize to a scenario in which all the photons have nearly the same energy so that we
can write,
\begin{equation}
\langle n_{\vec k} \rangle = \langle n_{\vec k'} \rangle \delta (k-k'),\label{e2}
\end{equation}
where $\langle n_{\vec k'} \rangle$ is given by (\ref{e1}), and $k \equiv
|\vec k|.$ Also $\vec k = \frac{2\pi \vec n}{L},$ where as usual $\vec n$ is a
vector with components, $0,\pm 1, \pm 2.$ The total number of photons $N$, in the volume $V$ being considered,
can be obtained in the usual way,
\begin{equation}
N = \frac{V}{(2\pi)^3} \int^\infty_o dk4\pi k^2 \langle n_k \rangle\label{e3}
\end{equation}
where $V$ is large. Inserting (\ref{e2}) in (\ref{e3}) we get,
\begin{equation}
N = \frac{2V}{(2\pi)^3} 4\pi k'^2 [\epsilon^\theta - 1]^{-1}[k], \theta
\equiv \beta \hbar \omega, \beta \equiv \frac{1}{KT}\label{e4},
\end{equation}
where $[k]$ is a dimensionality constant, introduced to compensate the loss of
a factor $k$ in the integral (\ref{e3}), owing to the $\delta$-function in
(\ref{e2})\\
We observe that, $\theta = \hbar \omega/KT \approx 1,$ since by (\ref{e2}), the
photons have nearly the same energy $\hbar \omega$. We also introduce,
\begin{equation}
\upsilon = \frac{V}{N}, \lambda = \frac{2\pi c}{\omega} = \frac{2\pi}{k}
\quad \mbox{and}\quad z = \frac{\lambda^3}{\upsilon}\label{e5}
\end{equation}
$\lambda$ being the wave length of the radiation. We now have from (\ref{e4}),
using (\ref{e5}),
\begin{equation}
z = \frac{8\pi}{k'(e-1)} =  \frac{8\pi c}{\omega (e-1)} = \frac{4\lambda}{(e-1)}[k]\label{e6}
\end{equation}
From (\ref{e6}) using the fact that $\hbar \omega \approx KT$, as seen
above, we conclude that, when
\begin{equation}
T = \frac{\hbar \omega}{K} \approx 3^\circ K\label{e7}
\end{equation}
then,
\begin{equation}
z \approx 1\label{e8}
\end{equation}
or conversely. As can be seen from (\ref{e6}), this corresponds to
\begin{equation}
\lambda \approx 0.4cm\label{e9}
\end{equation}
Remembering that from (\ref{e5}), $\lambda$ is the wave length and $v$
is the average volume per photon, the condition
(\ref{e8}) implies that all the photons are very densely packed as in the
case of Bose condensation. This means that from (\ref{e7}) and (\ref{e9}), we conclude
that at the background temperature of $3^\circ K$ or at the wave length $0.4 cm,$ in the micro-wave region, the radiation
"condenses" or has a peak intensity or conversely. The cosmic background radiation has the
maximum intensity exactly at the wave length (\ref{e9})\cite{r3}.\\
In a cosmological context, this result is valid for example in a quasi stationary
situation when approximate theremodynamical equilibrium is meaningful.
So is the observed Cosmic Background Radiation a result of photons settling
down to nearly the same energy and condensing at the temperature of
about $3^\circ K.$?\\
It may be mentioned, finally, that non-Big Bang explanations for this background
radiation have also been given\cite{r4,r5}.

\end{document}